\documentclass[twocolumn,showpacs,amssymb]{revtex4}

\usepackage{graphicx}
\usepackage{psfig}
\usepackage{dcolumn}
\usepackage{bm}

\newcommand{\ttherm}{$\tau_{\rm th}\ $}
\newcommand{\bag}{\mathcal{B}}
\def\lessim{\lower.5ex\hbox{$\; \buildrel < \over \sim \;$}}
\def\gtrsim{\lower.5ex\hbox{$\; \buildrel > \over \sim \;$}}

\begin{document} 
\topmargin -0.8cm
\preprint{}

\title{Hadronization of the Quark Universe}

\author{Michael J. Fromerth}
\author{Johann Rafelski}
\affiliation{Department of Physics, University of Arizona, Tucson, Arizona, 85721, USA}

\date{November 14, 2002; April 7, 2003}
\begin{abstract}
We explore quantitatively the conditions in which matter (protons, neutrons)
formed in the Early Universe  during
a period which spanned $10$--$50\, \mu$s. 
We obtain all chemical potentials implied by the present 
day baryon-to-photon  ratio. The baryochemical potential reaches
its minimal value at the beginning of hadronization of the quark Universe, 
where $\mu_b=1.1 \pm 0.25 $\,eV. In  the mixed hadron-quark phase, 
a significant hadron sector  electric charge distillation is found.
\end{abstract}

\maketitle

In the standard big-bang model, the large primordial baryon and antibaryon
abundance formed at hadronization of the deconfined quark-gluon plasma (QGP)
disappears due to mutual annihilation, exposing a slight net baryon 
number observed today \cite{Kolb90}. Applying the knowledge of equations 
of state of hadronic matter derived from the study of high energy
nuclear collisions \cite{Letessier02} we consider quantitatively this
evolution epoch of the early  Universe.

Our objective is to  quantify the values of the chemical potentials  
required to generate the observed  matter-antimatter asymmetry, and to use 
this to  quantify the magnitude of electrical charge distillation occurring 
during hadronization. While we do not enter here into the exploration 
of the consequences of these two findings, we note that
the values of the chemical potentials allow detailed study of hadron and lepton
 abundances during the epoch the universe cools from the hadronization
temperature to nucleosynthesis temperature. The dynamical process of
distillation can at QGP hadronization enhance an initially small 
baryon-antibaryon asymmetry to the value we encounter in our region of 
the Universe. 

The annihilation period 
began after the phase transformation from the QGP
 to a hot hadronic gas (HG), roughly $20$--$30\,\mu$s after the big bang 
(when the Universe was at a temperature of $\sim 170$~MeV), and 
continued until the density had diminished to the level that 
a ``nucleon freeze-out'' occurred at approximately 1\,ms ($T \simeq 35$~MeV). 
In this scenario,
annihilation of antimatter was very complete, quantified 
by the result  that the energy fraction in baryons and 
antibaryons in the  Universe dropped from $\sim 10\%$ when Universe 
was about 40\,$\mu$s old to $\sim 10^{-7}$ when it was one second 
old. 

The observational evidence about the antimatter non-abundance 
in the Universe is supported by the highly homogeneous
cosmic microwave background derived from the period of photon 
decoupling~\cite{Fixsen96}. This has been used  to 
argue that the  matter-antimatter domains on a scale smaller 
than the observable Universe are unlikely~\cite{Cohen98}; 
others see need for further experimental study to confirm 
this result~\cite{Kinney97}. 

The current small value of the baryon-to-photon ratio is the result of this
near complete annihilation of the large matter-antimatter abundance. 
Other than a (relatively) small increase in photons during nucleosynthesis 
and electron-ion recombination, the baryon and photon numbers should be 
preserved back to the period of annihilation.
Considering several observables, a range of $\eta$ is established and we
use here the latest WMAP result \cite{Spergel03}, 
$\eta \equiv n_{\rm B}/n_\gamma = 6.1^{+0.3}_{-0.2} \times 10^{-10}$.

The importance of $\eta$ is that it allows to determine the  
value of entropy  per baryon $S/B$ in the Universe,  which is 
conserved in adiabatic evolution. At present the entropy is 
dominated by photons,  and  nearly massless (decoupled) neutrinos.
It is straightforward to compute the entropy densities of 
these species from the partition function, and then to convert 
$\eta$ to $S/B$ using the  photon number density.
We obtain a value of $S/B = 8.0/\eta=1.3 \pm 0.1 \times 10^{10}$,
assuming a lower neutrino than photon temperature
(photons are reheated by $e^+e^-$ annihilation), and counting
only left/right-handed  neutrinos/antineutrinos.

{\bf Equations of State} 
We argue that the Universe was in chemical and thermal equilibrium 
around the time of hadronization, {\it i.e.}, when the quark-gluon
 Universe turned into a hadronic Universe.
To compute the thermodynamic properties of the QGP and HG phases, 
we study the partition functions $\ln{Z_{\rm QGP}}$ and $\ln{Z_{\rm HG}}$ 
as described in Ref.~\cite{Letessier02}:  we employ latest 
 QGP equations of state~\cite{Letessier03}, 
which model in detail with properties of quantum chromodynamics (QCD) 
at finite temperature and finite baryon density obtained in lattice QCD.
This approach involves quantum gases of quarks and effectively 
massive gluons with perturbative QCD corrections applied,
and a confining vacuum energy-pressure component 
$\bag = 258$~MeV~fm$^{-3}$.
In the HG partition function, we sum partial gas contributions 
including all hadrons  from Ref.~\cite{PDG98} having mass less 
than 2~GeV, and apply finite volume corrections \cite{Hagedorn80}.

Our use of partition functions assumes that local 
thermodynamic equilibrium (LTE) exists.
Considering the particle  spectra and yields 
measured at  the Relativistic Heavy Ion Collider at 
Brookhaven National Laboratory (RHIC), 
it is observed that a thermalization timescale 
on the order of \ttherm$\lessim 10^{-23}$~s is present 
in the QGP at hadronization~\cite{Letessier02}.
The mechanisms for such a rapid 
thermalization are at present under investigation. 
We expect this result to be valid qualitatively
in the primordial  QGP phase of matter. This then
assures us of LTE being present in the evolving Universe.
The local chemical equilibrium (LCE) is also approached 
at RHIC, indicating that this condition also prevails
in the early Universe. 

To apply these experimental results, we recall that 
the size of a  flat $k = 0$ Universe  
evolves as $R \propto t^{1/2}$, where $R$ is the size  
scale factor of the Universe~\cite{Battaner96}. 
This scaling is valid if the energy and 
momentum are dominated by radiation.
Furthermore, if the expansion is adiabatic 
and energy conserving, then 
$R \propto T^{-1}$.
The thermalization timescale is roughly 
$\tau_{\rm th}\approx  
{1}/{n\, \sigma\, v}$,
with $n$ the particle number density, $\sigma$ the 
cross section for (energy-exchanging) 
interactions, and $v$ the mean velocity.
Allowing for the change in relative velocity and a reduction in density, 
we can expect an increase in \ttherm as we cross from 
the relativistic QGP to the HG phase having strong non-relativistic components,
such that \ttherm$\lessim 10^{-22}$~s for the HG at $T = 170$~MeV. 
For a roughly constant value of $\sigma v$ during the expansion of the HG,
the thermalization timescale increases to \ttherm$\lessim 10^{-15}$~s at 
$T=1$~MeV.
At this point, the Universe is already one second old, so our 
assumption of LTE (and also of LCE) has a large margin of error 
and is in our opinion 
fully justified throughout the period of interest.

Chemical equilibration timescales are  significantly 
longer, due to smaller cross sections, 
than thermalization timescales. 
When chemical equilibrium cannot be  maintained in an expanding Universe, 
we find particle yield freeze-out. Near the phase
transformation from HG to QGP, chemical equilibrium for 
hadrons made of $u,d,s$ quarks is established. 
Hadron abundance evolution 
and deviations from the local equilibrium 
have not yet been studied in great detail
in the early Universe. Our estimates suggest that the
abundance of hadrons can be seen as being fixed by LCE
down to a few MeV. Note that annihilation depletion of
baryon density ceases  around temperature $T\simeq 35$\,MeV. 

In a system of non-interacting particles, the chemical 
potential $\mu_i$ of each species $i$ is independent of 
the chemical potentials of other species, resulting 
in a large number of free parameters.
The many chemical particle interactions occurring in the QGP and HG phases,
 however, greatly reduce this number.

First, in thermal equilibrium, photons assume the Planck 
distribution, implying a zero photon chemical potential; i.e., $\mu_\gamma = 0$.
Next, for any reaction $\nu_i A_i = 0$, where $\nu_i$ are the 
reaction equation coefficients of the chemical species $A_i$, 
chemical equilibrium occurs when $\nu_i \mu_i = 0$, which 
follows from a minimization of the Gibbs free energy.
Because reactions such as $f + \bar{f} \rightleftharpoons 2 \gamma$ 
are allowed, where $f$ and $\bar{f}$ are a fermion -- antifermion 
pair, we immediately see that $\mu_f = -\mu_{\bar{f}}$ whenever 
chemical and thermal equilibrium have been attained.

Furthermore, when the system is chemically equilibrated 
with respect to weak interactions, we can write down the 
following relationships~\cite{Glendenning00}:
\begin{equation}
\mu_e - \mu_{\nu_e}=\mu_\mu - \mu_{\nu_{\mu}}=\mu_{\tau} - \mu_{\nu_{\tau}}\ 
\equiv\ \Delta \mu_l,
\end{equation}
\begin{equation}
\mu_u=\mu_d - \Delta \mu_l,\qquad \mu_s=\mu_d \ ,
\end{equation}
with the chemical equilibrium of hadrons being equal to the 
sum of the chemical potentials of their constituent quarks; 
{\it e.g.} for $\, \Sigma^0\,$({\it uds}):
 $\mu_{\Sigma^0}=\mu_u + \mu_d + \mu_s=3\, \mu_d - \Delta \mu_l$, and the baryochemical potential is:
\begin{equation}
\mu_b=\frac32(\mu_d +\mu_u)=3 \mu_d -\frac32 \Delta \mu_l.
\end{equation}
Finally, if the experimentally-favored ``large mixing angle'' solution~\cite{Ahmad02} is correct, the neutrino oscillations $\nu_e \rightleftharpoons \nu_\mu \rightleftharpoons \nu_\tau$ imply that~\cite{Raffelt02}:
$
\mu_{\nu_e} = \mu_{\nu_{\mu}} =\mu_{\nu_{\tau}} \equiv \mu_\nu,
$
which reduces the number of independent chemical potentials to three.
We choose these to be $\mu_d$, $\mu_e$, and $\mu_\nu$. 
We next seek to establish what values of our independent 
chemical potentials are required to generate the observed 
matter-antimatter asymmetry.

{\bf Single Phase} 
In a homogeneous (i.e., single phase) 
Universe, we seek to satisfy the following three criteria:\\
\noindent {\it i) Charge neutrality} ($Q = 0$) is required 
to eliminate Coulomb energy.  This implies that:
\begin{equation}
n_Q\equiv \sum_i\, Q_i\, n_i (\mu_i, T)=0, 
\label{Q0}
\end{equation}
where $Q_i$ and $n_i$ are the charge and number density of species $i$, 
and the summation is carried out over all species present in the phase.\\
\noindent {\it ii)  Net lepton number equals net baryon number} 
($L = B$) is required in many baryogenesis scenarios.  This implies that:
\begin{equation}
n_L - n_B\equiv \sum_i\, (L_i - B_i)\, n_i (\mu_i, T)=0 ,
\label{LB0}
\end{equation}
where $L_i$ and $B_i$ are the lepton and baryon numbers of species $i$.\\
\noindent {\it iii)  Constant entropy-per-baryon} ($S/B$).  
This is the statement that the Universe evolves adiabatically, 
and is equivalent to:
\begin{equation}
\frac{s}{n_B}\equiv\frac{\sum_i\, s_i(\mu_i, T)}{\sum_i\, B_i\, n_i(\mu_i, T)}
=1.3 \pm 0.1 \times 10^{10}, 
\label{SperB}
\end{equation}
where $s_i$ is the entropy density of species $i$.

\begin{figure}[t]
\resizebox{3.25in}{!}{\rotatebox{0}{\includegraphics{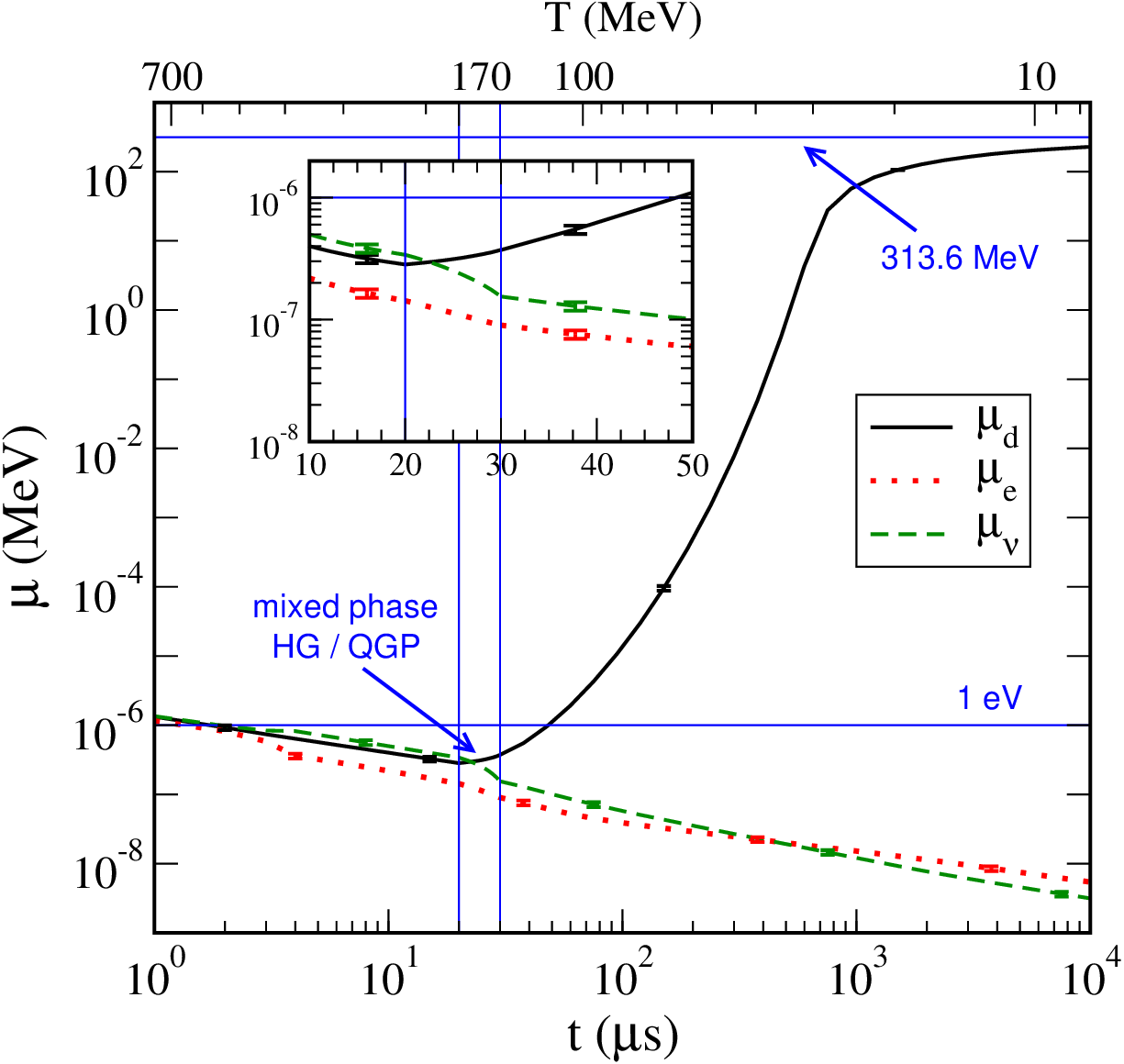}}}
\caption{Chemical potentials $\mu_d$, $\mu_e$, and $\mu_\nu$
around the time of the QGP-HG phase transformation.  
The error bars arise from the uncertainty in $\eta\,$.  
Insert --- expanded view around the phase transformation.
Horizontal and vertical lines inserted to guide the eye.}
\label{chem_pot}
\end{figure}

Eqs.~(\ref{Q0})--(\ref{SperB}) 
constitute a system of three coupled, non-linear equations 
of three unknowns ($\mu_d$, $\mu_e$, and $\mu_\nu$) 
for a given temperature.
These equations were solved numerically using the 
Levenberg-Marquardt method~\cite{Press92} to obtain 
Fig.~\ref{chem_pot}, which shows the values of 
$\mu_d$, $\mu_e$, and $\mu_\nu$ as function of the age of 
the Universe (the top axis shows the corresponding temperature).
As the temperature decreases, the value of $\mu_d$ 
approaches weighted one-third of the nucleon mass 
($(2m_n-m_p)/3\simeq 313.6$~MeV, see horizontal 
line in Fig.~\ref{chem_pot}).
This follows because the baryon (proton 
and neutron) partition functions 
are in the classical Boltzmann regime at these temperatures.

The error bars arise from WMAP uncertainty in the value of $\eta$.
The chemical potentials required to generate 
the current matter-antimatter asymmetry are  smaller 
than 1~eV (horizontal line in Fig.~\ref{chem_pot}).
Near the Universe hadronization temperature $T_h$ 
additional sensitivity in the value of chemical 
potentials arises due to the uncertainty in the
value of hadronization temperature. With 6\% uncertainty  in 
$T_h=170\pm 10$ MeV we obtain 15\% uncertainty in the minimal 
value of the d-quark chemical potential. 
This error is dominating the uncertainty of the minimal value
of the baryochemical potential, and thus
$\mu_b|_{\rm min}=\simeq 1.1\pm 0.25$  eV.

\begin{figure}[t]
\resizebox{3.0in}{!}{\rotatebox{270}{\includegraphics{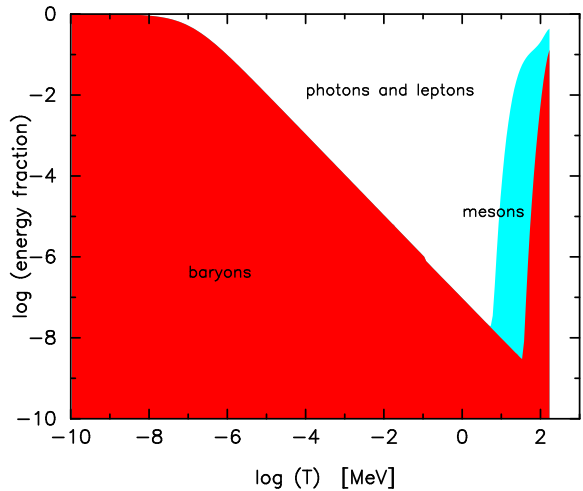}}}
\caption{The hadronic energy content of the luminous Universe as a function 
of temperature assuming a constant entropy-per-baryon number 
of $1.3 \times 10^{10}$.}
\label{e_frac}
\end{figure}

\begin{figure}[t]
\resizebox{3.0in}{!}{\rotatebox{270}{\includegraphics{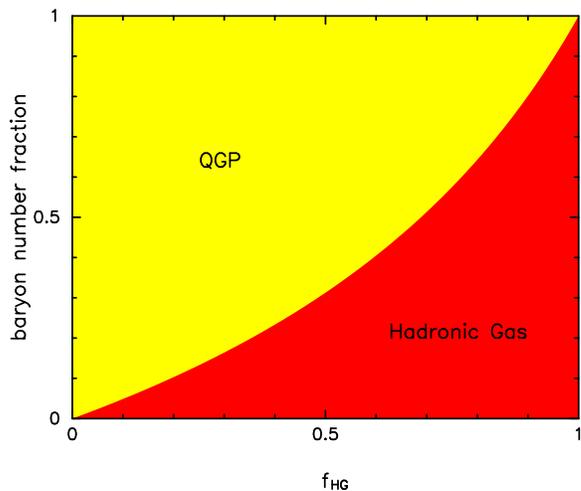}}}
\caption{The fraction of baryons in the HG and QGP during 
phase transformation. The parameter $f_{\rm HG}$ is the fraction 
of total phase space occupied by the hadronic gas phase.}
\label{bary_frac}
\end{figure}

\begin{figure}
\resizebox{2.75in}{!}{\rotatebox{0}{\includegraphics{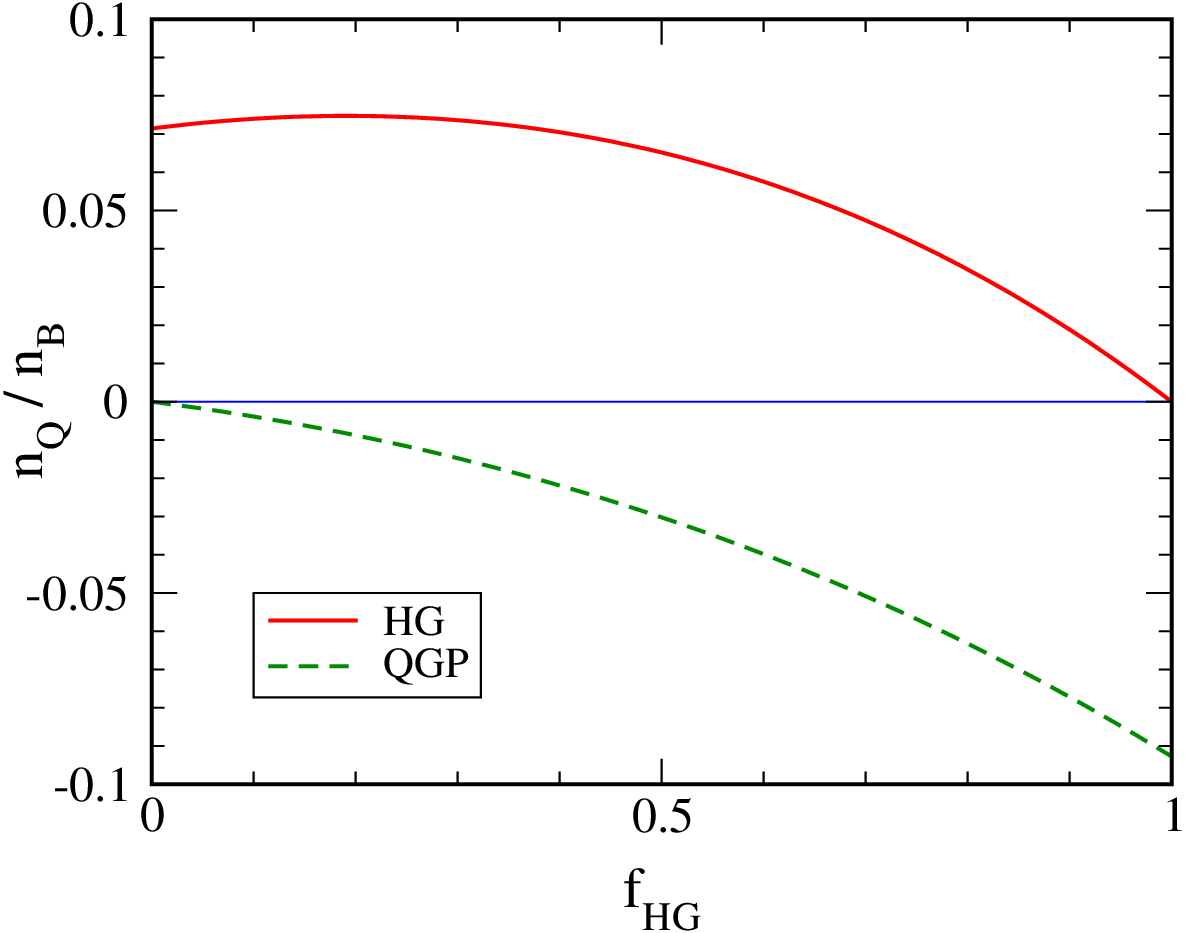}}}
\caption{Net charge (including leptons) per net baryon 
number in the HG and QGP during phase transformation.  Horizontal line at zero inserted to guide the eye.}
\label{QperB}
\end{figure}

Given the chemical potentials of all particles we can
trace out all particle densities in the early Universe and
can, for example, derive the hadronic energy content 
in the luminous Universe as a function of temperature, 
shown in Figure~\ref{e_frac}.
The fraction of energy in baryons and antibaryons 
is roughly 10\% at the QGP-HG phase transformation, 
but rapidly vanishes, becoming significant again only 
when the Universe has cooled, and enters
nucleon rest mass matter-dominated  era.

{\bf Phase Transformation:  QGP to HG} 
During the QGP to HG phase transformation, when 
both phases co-exist, the macroscopic conditions 
i.~--~iii.~above  are no longer valid individually within either 
the QGP or HG phases, and that the correct expressions must 
contain combinations of the two phases.
We therefore parameterize the partition function 
during the phase transformation as $\ln{Z_{\rm tot}}=f_{\rm HG} 
\ln{Z_{\rm HG}}+(1 - f_{\rm HG}) \ln{Z_{\rm QGP}}$,
where the factor $f_{\rm HG}$ represents the 
fraction of total phase space occupied by the HG phase.
The correct expression analogous to Eq.~(\ref{Q0}) is:
\begin{eqnarray}
Q =  0 
 & = & n_Q^{\rm QGP}\, V_{\rm QGP}+n_Q^{\rm HG}\, V_{\rm HG} \nonumber \\
 & = & V_{\rm tot} \left[ (1-f_{\rm HG})\, n_Q^{\rm QGP}+f_{\rm HG}\, n_Q^{\rm HG} \right],
\end{eqnarray}
where the total volume $V_{\rm tot}$ is irrelevant to the solution.
Analogous expressions can be derived for Eqs.\,(\ref{LB0})~and~(\ref{SperB}).

These expressions were used to obtain Fig.~\ref{bary_frac}, 
which shows the fraction of the total baryon number in the QGP 
and HG phases as a function of $f_{\rm HG}$. We note that
the QGP phase has a higher baryon number density 
than the HG phase throughout the transformation.
This value is nearly constant, with 
$n_B^{\rm QGP} / n_B^{\rm HG} \approx 3$.

In Fig.~\ref{chem_pot}, it was assumed that the value of 
$f_{\rm HG}$ evolved linearly in time and that the duration 
of the phase transformation was $\tau_h=10\,\mu$s.
In reality, these quantities are sensitive to the properties
of the equations of state and the dynamics of the phase transformation.
A more complete evaluation will  require application of transport theory and
is beyond the scope of the current work. Our value of $\tau_h$ is an estimate
discussed in~\cite{Letessier02}.

Figure~\ref{QperB} shows the net charge per baryon in 
each phase as a function of $f_{\rm HG}$.
At its onset the small region of HG phase takes on an initial positive 
charge density, which can be attributed to the proton-neutron 
bias toward positive charge.
As a result, the QGP domain takes on a (initially tiny) negative charge density.
Such distilled dynamical asymmetry in particle yields was previously explored for strangeness 
separation and associated strangelet formation~\cite{Greiner87,Greiner91}.

Since the sign of the effect seen in Figure~\ref{QperB}
is the same across the entire hadronization region, the total charge of the remaining QGP domains is ever-increasingly 
negative and one would expect development of electromagnetic potential,
which effectively alters the values of chemical potentials for charged species.
It is evident  that the process of charge distillation will have a feed-back 
effect on the QGP-HG transformation, and that flows of particles will
occur that will alter the uniformly small net baryon 
density~\cite{Witten84}. This can
affect (during the phase transformation) any local initial baryon-antibaryon asymmetry, and may also serve as a mechanism for generating 
 magnetic fields in the primordial Universe~\cite{Cheng94}. Quantitative
evaluation of this baryon asymmetry enhancement effect entails
methods of advanced transport theory beyond the scope of this work. 

We have determined the chemical potentials required to describe
the matter-antimatter asymmetry in the Early Universe.
We have shown that the non-zero chemical potentials result in 
a charge distillation during the phase transformation, with the QGP 
and HG receiving  negative and positive charge densities, respectively.
We note that a  separation of baryons and antibaryons into domains 
could maintain a homogeneous zero charge density Universe, a 
phenomenon which could, {\it e.g.}, play a significant role in amplifying a 
pre-existent, much smaller net baryon yield, e.g. arising within 
the realm of the standard model. The evolution of the chemical
potentials allows to trace out the particle abundance in the 
early universe prior to nucleosynthesis.   

\vskip 0.3cm
\noindent{\bf Acknowledgments:}
Work supported in part by a grant from the U.S. Department of
Energy,  DE-FG03-95ER40937. 


\begin{thebibliography}{19}
\expandafter\ifx\csname natexlab\endcsname\relax\def\natexlab#1{#1}\fi
\expandafter\ifx\csname bibnamefont\endcsname\relax
  \def\bibnamefont#1{#1}\fi
\expandafter\ifx\csname bibfnamefont\endcsname\relax
  \def\bibfnamefont#1{#1}\fi
\expandafter\ifx\csname citenamefont\endcsname\relax
  \def\citenamefont#1{#1}\fi
\expandafter\ifx\csname url\endcsname\relax
  \def\url#1{\texttt{#1}}\fi
\expandafter\ifx\csname urlprefix\endcsname\relax\def\urlprefix{URL }\fi
\providecommand{\bibinfo}[2]{#2}
\providecommand{\eprint}[2][]{\url{#2}}

\bibitem[{\citenamefont{{Kolb} and {Turner}}(1990)}]{Kolb90}
\bibinfo{author}{\bibfnamefont{E.~W.} \bibnamefont{{Kolb}}} \bibnamefont{and}
  \bibinfo{author}{\bibfnamefont{M.~S.} \bibnamefont{{Turner}}},
  \emph{\bibinfo{title}{The Early Universe}}
  (\bibinfo{publisher}{Addison-Wesley}, \bibinfo{year}{1990}).

\bibitem[{\citenamefont{{Letessier} and {Rafelski}}(2002)}]{Letessier02}
\bibinfo{author}{\bibfnamefont{J.}~\bibnamefont{{Letessier}}} \bibnamefont{and}
  \bibinfo{author}{\bibfnamefont{J.}~\bibnamefont{{Rafelski}}},
  \emph{\bibinfo{title}{Hadrons and Quark-Gluon Plasma}}
  (\bibinfo{publisher}{Cambridge}, \bibinfo{year}{2002}).

\bibitem[{\citenamefont{{Fixsen} et~al.}(1996)\citenamefont{{Fixsen}, {Cheng},
  {Gales}, {Mather}, {Shafer}, and {Wright}}}]{Fixsen96}
\bibinfo{author}{\bibfnamefont{D.~J.} \bibnamefont{{Fixsen}}},
  \bibinfo{author}{\bibfnamefont{E.~S.} \bibnamefont{{Cheng}}},
  \bibinfo{author}{\bibfnamefont{J.~M.} \bibnamefont{{Gales}}},
  \bibinfo{author}{\bibfnamefont{J.~C.} \bibnamefont{{Mather}}},
  \bibinfo{author}{\bibfnamefont{R.~A.} \bibnamefont{{Shafer}}},
  \bibnamefont{and} \bibinfo{author}{\bibfnamefont{E.~L.}
  \bibnamefont{{Wright}}}, \bibinfo{journal}{Astrophys. J.}
  \textbf{\bibinfo{volume}{473}}, \bibinfo{pages}{576} (\bibinfo{year}{1996}).

\bibitem[{\citenamefont{{Cohen} et~al.}(1998)\citenamefont{{Cohen}, {de
  Rujula}, and {Glashow}}}]{Cohen98}
\bibinfo{author}{\bibfnamefont{A.~G.} \bibnamefont{{Cohen}}},
  \bibinfo{author}{\bibfnamefont{A.}~\bibnamefont{{de Rujula}}},
  \bibnamefont{and} \bibinfo{author}{\bibfnamefont{S.~L.}
  \bibnamefont{{Glashow}}}, \bibinfo{journal}{Astrophys. J.}
  \textbf{\bibinfo{volume}{495}}, \bibinfo{pages}{539} (\bibinfo{year}{1998}).

\bibitem[{\citenamefont{{Kinney} et~al.}(1997)\citenamefont{{Kinney}, {Kolb},
  and {Turner}}}]{Kinney97}
\bibinfo{author}{\bibfnamefont{W.~H.} \bibnamefont{{Kinney}}},
  \bibinfo{author}{\bibfnamefont{E.~W.} \bibnamefont{{Kolb}}},
  \bibnamefont{and} \bibinfo{author}{\bibfnamefont{M.~S.}
  \bibnamefont{{Turner}}}, \bibinfo{journal}{Phys. Rev. Lett.}
  \textbf{\bibinfo{volume}{79}}, \bibinfo{pages}{2620} (\bibinfo{year}{1997}).

\bibitem[{\citenamefont{{Spergel} et~al.}(2004)}]{Spergel03}
\bibinfo{author}{\bibfnamefont{D.}~\bibnamefont{{Spergel}}} 
\bibnamefont{et~al.},
  \bibinfo{journal}{astro-ph/0302209}  (\bibinfo{year}{2003}).


\bibitem[{\citenamefont{{Letessier} and {Rafelski}}(2003)}]{Letessier03}
\bibinfo{author}{\bibfnamefont{J.}~\bibnamefont{{Letessier}}} \bibnamefont{and}
  \bibinfo{author}{\bibfnamefont{J.}~\bibnamefont{{Rafelski}}},
  \bibinfo{journal}{Phys. Rev. C} \textbf{\bibinfo{volume}{67}},
  \bibinfo{pages}{031902R} (\bibinfo{year}{2003}).
\bibinfo{author}{\bibfnamefont{S.}~\bibnamefont{{Hamieh}}},
  \bibinfo{author}{\bibfnamefont{J.}~\bibnamefont{{Letessier}}},
  \bibnamefont{and}
  \bibinfo{author}{\bibfnamefont{J.}~\bibnamefont{{Rafelski}}},
  \bibinfo{journal}{Phys. Rev. C} \textbf{\bibinfo{volume}{62}},
  \bibinfo{pages}{064901} (\bibinfo{year}{2000}).

%

\bibitem[{\citenamefont{{Caso} et~al.}(1998)}]{PDG98}
\bibinfo{author}{\bibfnamefont{C.}~\bibnamefont{{Caso}}} \bibnamefont{et~al.},
  \bibinfo{journal}{Eur. Phys. J. C} \textbf{\bibinfo{volume}{3}},
  \bibinfo{pages}{1} (\bibinfo{year}{1998}).

\bibitem[{\citenamefont{{Hagedorn} and {Rafelski}}(1980)}]{Hagedorn80}
\bibinfo{author}{\bibfnamefont{R.}~\bibnamefont{{Hagedorn}}} \bibnamefont{and}
  \bibinfo{author}{\bibfnamefont{J.}~\bibnamefont{{Rafelski}}},
  \bibinfo{journal}{Phys. Lett. B} \textbf{\bibinfo{volume}{97}},
  \bibinfo{pages}{136} (\bibinfo{year}{1980}).

\bibitem[{\citenamefont{{Battaner}}(1996)}]{Battaner96}
\bibinfo{author}{\bibfnamefont{E.}~\bibnamefont{{Battaner}}},
  \emph{\bibinfo{title}{{Astrophysical Fluid Dynamics}}}
  (\bibinfo{publisher}{Cambridge}, \bibinfo{year}{1996}).

\bibitem[{\citenamefont{{Glendenning}}(2000)}]{Glendenning00}
\bibinfo{author}{\bibfnamefont{N.~K.} \bibnamefont{{Glendenning}}},
  \emph{\bibinfo{title}{Compact Stars: Nuclear Physics, Particle Physics, and
  General Relativity}} (\bibinfo{publisher}{Springer}, \bibinfo{year}{2000}).

\bibitem[{\citenamefont{{Ahmad} et~al.}(2002)}]{Ahmad02}
\bibinfo{author}{\bibfnamefont{Q.~R.} \bibnamefont{{Ahmad}}}
  \bibnamefont{et~al.}, \bibinfo{journal}{Phys. Rev. Lett.}
  \textbf{\bibinfo{volume}{89}}, \bibinfo{pages}{11302} (\bibinfo{year}{2002}).

\bibitem[{\citenamefont{{Raffelt}}(2002)}]{Raffelt02}
\bibinfo{author}{\bibfnamefont{G.~G.} \bibnamefont{{Raffelt}}},
  \bibinfo{journal}{hep-ph/0208024}  (\bibinfo{year}{2002});
\bibnamefont{and}:
  \bibinfo{author}{\bibfnamefont{A.D.}~\bibnamefont{Dolgov}}, 
  \bibinfo{author}{\bibfnamefont{S.H.}~\bibnamefont{Hansen}}, 
  \bibinfo{author}{\bibfnamefont{S.}~\bibnamefont{Pastor}}, 
  \bibinfo{author}{\bibfnamefont{S.T.}~\bibnamefont{Petcov}}, 
  \bibinfo{author}{\bibfnamefont{G.G.}~\bibnamefont{Raffelt}}, 
\bibnamefont{and}
  \bibinfo{author}{\bibfnamefont{D.V.}~\bibnamefont{Semikoz}}, 
  \bibinfo{journal}{Nucl. Phys.} \textbf{\bibinfo{volume}{B632}},
  \bibinfo{pages}{363} (\bibinfo{year}{2002}).


\bibitem[{\citenamefont{{Press} et~al.}(1992)\citenamefont{{Press},
  {Teukolsky}, {Vetterling}, and {Flannery}}}]{Press92}
\bibinfo{author}{\bibfnamefont{W.~H.} \bibnamefont{{Press}}},
  \bibinfo{author}{\bibfnamefont{S.~A.} \bibnamefont{{Teukolsky}}},
  \bibinfo{author}{\bibfnamefont{W.~T.} \bibnamefont{{Vetterling}}},
  \bibnamefont{and} \bibinfo{author}{\bibfnamefont{B.~P.}
  \bibnamefont{{Flannery}}}, \emph{\bibinfo{title}{{Numerical recipes in C. The
  art of scientific computing}}} (\bibinfo{publisher}{Cambridge},
  \bibinfo{year}{1992}).

\bibitem[{\citenamefont{Greiner et~al.}(1987)\citenamefont{Greiner, Koch, and
  Stocker}}]{Greiner87}
\bibinfo{author}{\bibfnamefont{C.}~\bibnamefont{Greiner}},
  \bibinfo{author}{\bibfnamefont{P.}~\bibnamefont{Koch}}, \bibnamefont{and}
  \bibinfo{author}{\bibfnamefont{H.}~\bibnamefont{Stocker}},
  \bibinfo{journal}{Phys. Rev. Lett.} \textbf{\bibinfo{volume}{58}},
  \bibinfo{pages}{1825} (\bibinfo{year}{1987}).

\bibitem[{\citenamefont{Greiner and Stocker}(1991)}]{Greiner91}
\bibinfo{author}{\bibfnamefont{C.}~\bibnamefont{Greiner}} \bibnamefont{and}
  \bibinfo{author}{\bibfnamefont{H.}~\bibnamefont{Stocker}},
  \bibinfo{journal}{Phys. Rev.} \textbf{\bibinfo{volume}{D44}},
  \bibinfo{pages}{3517} (\bibinfo{year}{1991}).

\bibitem[{\citenamefont{Witten}(1984)}]{Witten84}
\bibinfo{author}{\bibfnamefont{E.}~\bibnamefont{{Witten}}},
  \bibinfo{journal}{Phys. Rev. D} \textbf{\bibinfo{volume}{30}},
  \bibinfo{pages}{272} (\bibinfo{year}{1984});

\bibitem[{\citenamefont{{Cheng} and {Olinto}}(1994)}]{Cheng94}
\bibinfo{author}{\bibfnamefont{B.}~\bibnamefont{{Cheng}}} \bibnamefont{and}
  \bibinfo{author}{\bibfnamefont{A.~V.} \bibnamefont{{Olinto}}},
  \bibinfo{journal}{Phys. Rev. D} \textbf{\bibinfo{volume}{50}},
  \bibinfo{pages}{2421} (\bibinfo{year}{1994});
\bibnamefont{and}:
\bibinfo{author}{\bibfnamefont{G.}~\bibnamefont{{ Sigl}}}, 
  \bibinfo{author}{\bibfnamefont{A.~V.} \bibnamefont{{Olinto}}},
\bibnamefont{and}
  \bibinfo{author}{\bibfnamefont{K.} \bibnamefont{{Jedamzik}}},
  \bibinfo{journal}{Phys. Rev. D} \textbf{\bibinfo{volume}{55}},
  \bibinfo{pages}{4582} (\bibinfo{year}{1997}).


\end{thebibliography}

\end{document}